\documentclass[twocolumn,showpacs,preprintnumbers,amsmath,amssymb]{revtex4}

\usepackage{graphicx}
\usepackage{dcolumn}
\usepackage{bm}

\begin{document}

\preprint{APS/123-QED}

\title{Spatio temporal characterization of interfacial Faraday
waves:\\ A new absorption technique}

\author{A.V.~Kityk}
\affiliation{Institute for Computer Science, Technical University
of Czestochowa, Electrical Eng. Dep., Al. Armii Krajowej 17,
PL-42200 Czestochowa, Poland}
\author{J.Embs, V. V. Mekhonoshin}
\affiliation{Theoretische Physik, Universit\"at des Saarlandes,
66041 Saarbr\"ucken, Germany}
\author{C. Wagner}
 \email{c.wagner@mx.uni-saarland.de}
\affiliation{Experimentalphysik, Universit\"at des Saarlandes,
66041 Saarbr\"ucken, Germany}

\date{\today}

\begin{abstract}
We present measurements of the complete spatio-temporal Fourier
spectrum of Faraday waves. The Faraday waves are generated at the
interface of two immiscible index matched liquids of different
density. By use of a new absorption technique we are able to
determine the bifurcation scenario from the flat surface to the
patterned state for each complex spatial and temporal Fourier
component separately. The measured surface spectra at onset are in
good agreement with the predictions from a linear stability
analysis. For the nonlinear state our measurements show in a
direct manner how energy is transferred from lower to higher
harmonics and we quantify the nonlinear coupling coefficients.
Furthermore we find that the nonlinear coupling generates constant
components in the spatial Fourier spectrum leading to a
contribution of a non oscillating permanent sinusoidal deformed
surface state. A comparison of a hexagonal and a rectangular
pattern reveals that spatial resonances only can give rise to a
spectrum that violates the temporal resonance conditions given by
the weakly nonlinear theory.
\end{abstract}


\maketitle

\section*{1. Introduction}
The Faraday Experiment is probably the first non equilibrium
pattern forming system that has been investigated scientifically,
namely by Michael Faraday in 1831 \cite{Faraday}. Nevertheless it
was only recently that it was possible to determine the complete
Fourier spectrum of the deformed surface state \cite{Kityk04}.
While an experimental analysis of the full mode spectrum in other
pattern forming model systems like Rayleigh-Benard or
Taylor-Couette has been standard technique for a long time, it is
the refraction of light at the free surface of a liquid that
renders the analysis of surface waves so difficult. Quantitative
information about the patterned state up to higher orders is
important not only to verify the validity of theoretical
calculations \cite{kumar94,zhang96} but also to gain insight on
the resonance mechanisms that \textit{form} the patterns. The
Faraday experiment is especially known for its richness of
different patterns that are observed
\cite{miles90,milner90,Christiansen92,edwards94,binks97,kudrolli97}.
By using complex liquids \cite{wagner99}, very low fill heights
\cite{wagner00}  or by introducing additional driving frequencies,
highly complex ordered states like superlattices
\cite{kudrolli98,arbell98} have been observed recently. But we
will demonstrate that even simple patterns like lines, squares and
hexagons observed in a single driving frequency experiment can
still unveil unknown surprising characteristics.

A discussion of the different attempts to reveal quantitative
information on the the surface elevation profile $\zeta(r,t)$ of
Faraday waves is given in \cite{Wernet01}. The main difficulties
in determining the surface elevation profile of capillary waves at
the the free surface are the difference in refractive indexes of
the liquid and the air, and the fact that the interface diffuses
almost no light but rather reflects or transmits incoming light
completely. To our knowledge there is only one optical method that
overcomes this problem with the use Polystyrene colloids to
provide light scatterers within the fluid \cite{Wright96}, but the
method was only used in a turbulent regime. Another powerful
method for the investigation of capillary waves on ferrofluids
based on x-ray absorption was presented in \cite{Richter01}, but
the related costs and efforts might be justified for fully opaque
liquids only. To bypass the problems associated with light
refraction and reflection on a liquid-air interface we chose to
study the interface between two index matched liquids. The upper
fluid is transparent, the  lower one is dyed. In the presence of
surface deformations the instantaneous thickness can be deduced
from the intensity of the light transmitted trough the colored
layer. From a hydrodynamic point of view the replacement of the
air by a second liquid is nothing but a change of viscosity and
density, though the low kinematic viscosity of air simplifies the
theoretical calculations. However, the first exact theoretical
analysis of the linear stability problem by Kumar and Tuckerman
\cite{kumar94} was carried out for the more general case of a
system of two layers of liquid.

\section*{2. Experimental setup}
\begin{figure}
\includegraphics[ width=0.5\linewidth]{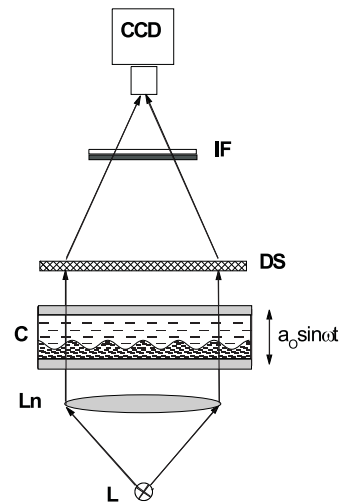}
 \caption{Experimental setup: L - halogen lamp, Ln -
lens, C - container filled by two liquids: SOIL and WSS with the
same refractive indices,  DS - diffusive screen, IF - interference
filter, CCD - high speed CCD camera. } \label{fig1}
\end{figure}

The experimental setup is shown in Fig.~\ref{fig1}.  The container
consists of an aluminium ring (diameter $D=$18 cm) seperating two
parallel glass windows by a gap of 10 mm. It is filled by two
unmixible liquids: a silicone oil (SOIL, Dow Corning, viscosity
$\eta = 20$ mPas, density $\rho=949$ kg/m$^3$) and an aqueous
solution of sugar and NiSO$_4$(WSS, $\eta = 7.2 $ mPas,
$\rho=1185$ kg/m$^3$). The liquid liquid interfacial tension has
been determined with the pending droplet method to $35 \pm 2$
dynes/cm. The ratio of the filling heights SOIL/WSS was 8.4/1.6.
The choice of heights was made in order to obtain a variety of
different patterns, including a transition from squares to
hexagons \cite{binks97b}. The sugar concentration has been adapted
to match the refractive index to that of the covering silicone oil
($n \simeq 1.405$) to a precision of $5 \times 10^{-4}$. The
Ni$^{2+}$-ions produce a broad absorbtion band in the spectral
region 600-800 nm and provide high contrast patterns projected
onto the diffusive screen. The container is illuminated from below
with parallel light, and a band pass filter in front of the camera
was used to detect only wavelengths $\lambda = (655 \pm 5) nm$. By
varying the intensity of the lamp the flat interface has been set
to a level of about $50\%$ of the maximum optical transmission. At
a NiSO$_4$ concentration of $17 \%$ by weight the contrast between
the light intensity passing through crests and valleys of the wave
pattern was optimum. The associated coefficient of optical
absorption was measured as $\alpha = 5.2 \pm 0.1 cm^{-1}$. In
order to avoid uncontrolled changes of the viscosity, density and
interfacial tension of both liquids all the measurements were
performed at a constant temperature (23$\pm$0.1$^o$C). The Faraday
waves were excited by an electromagnetic shaker vibrating
vertically with an acceleration in the form $a(t) =
a_0$cos$(\Omega t)$. The driving signal came from a computer via a
D/A-converter and the acceleration has been measured by
piezoelectric sensor. A self developed closed-loop algorithm was
used to suppress higher harmonics $n(\Omega t)$ in the driving
signal to guarantee a purely harmonic driving. Faraday patterns
were recorded in the following way: a high speed (250 Hz) 8-bit
CCD camera was mounted some distance above the diffusive screen.
Pictures were taken synchronous to the external driving. For a
certain instant $t_o$ the surface elevation of the Faraday
patterns $h (x,y,t_o)$ is given by:
\begin{equation}
h(x,y,t_o)= \frac {1}{\alpha}ln \frac{I_r(x,y)}{I_p(x,y,t_o)}
\label{eqn1}
\end{equation}
where $I_r(x,y)$ and $I_p(x,y,t_o)$ are 2D intensity distributions
captured by the camera for the reference picture (flat interface,
$a_o=0$) and for the Faraday pattern ($a_o \ne 0$), respectively.
Finally, the surface elevation function $h (x,y,t)$ are Fourier
transformed and the time evolution of the Fourier amplitudes and
phases of spatial modes is extracted. The use of a high speed
camera compared to the earlier measurements by some of the
co-authors \cite{Kityk04} allows for a better temporal resolution
and the method is not sensitive to distortions (defects) on time
scales of several periods. The logarithm of the intensity profile
renders the dynamic range nonlinear, and with an 8-bit dynamical
range the resolution is approximately $1\%$ ($2\%$) at small
(high) surface elevations, relative to the maximal surface
heights. The validity of the method has been also checked with
flat layers of colored liquids of different thicknesses and the
same accuracy was found. However, one should note the the Fourier
transformation integrates over many pixels and a significant
better resolution is to expect.

\section*{3. The linear regime}

\begin{figure}
\includegraphics[ width=0.95\linewidth]{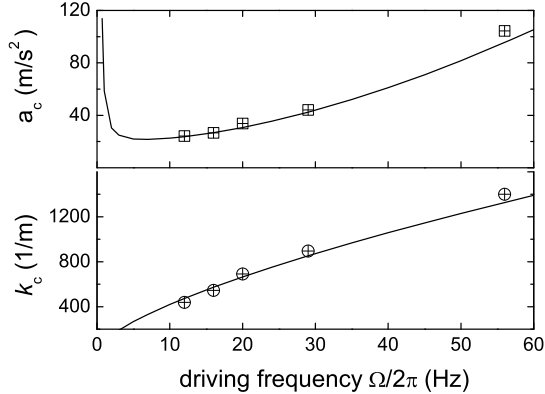}
\caption{Critical acceleration $a_c$ and critical wave number
$k_c$ for different driving frequencies $\Omega$. The symbols mark
experimental data, the lines the theoretical linear stability
analysis. The size of the symbols coincide with the size of the
error bars.}\label{lin1}
\end{figure}

The experiments have been performed by quasistatically ramping the
driving amplitude for the frequencies $f=\Omega/2\pi=12,16,20,29$
and $57 Hz$ from slightly below the critical acceleration $a_c$
($\varepsilon = (a-a_c)/a_c= -0.02)$ up to just below the
acceleration where the interface disintegrates and droplets form.
In the same form a ramp was driven down to check for hysteretic
effects, of which none were found. For each amplitude step a
series of pictures where taken and Fourier transformed. Typically
the pattern occurs in the center region of the container first but
evolves in a range of $\Delta\varepsilon = 0.02$. From the Fourier
transformation of the pictures at $-0.02<\varepsilon < 0.1$ (Fig.
\ref{squ}) the critical acceleration $a_c$ and the critical
wavenumber $k_c$ has been determined (Fig. \ref{lin1}).
The experimental data can be compared with the results from the
theoretical linear stability analysis that has been performed
using the algorithm proposed by Kumar and Tuckerman\cite{kumar94}.
The agreement between theory and experiment  is very good, similar
to former studies at the liquid-air interface
\cite{bechhoefer95,wagner97}. But with our new technique we are
now able to verify more details of the predictions of the linear
theory, e.g. the temporal spectra at onset of the instability. It
is a particular feature of the Faraday-Experiment, that at onset
only one wave number $k_c$ becomes unstable, but the temporal
spectrum already contains  multiples of the fundamental
oscillation frequency $\omega$ at onset. We are in the regime of
subharmonic response and the fundamental oscillation frequency at
onset is always $\omega=\Omega/2$ but the spectrum also contains
$(n+1/2) \Omega$ frequency components. More precisely we can write
the surface deformation $h({\bf r,t})$ as
\begin{equation}
h({\bf r}, t)= \frac{1}{4} \, \sum_{i=1}^{N} {(A_i e^{{\rm i} {\bf
k}_i \cdot {\bf r}} } + c.c.) \, \sum_{n=-\infty}^{+\infty}
\zeta_n e^{{\rm i} (n+1/2) \Omega t} \label{eqn2}
\end{equation}
Here ${\bf r}=(x,y)$ are the horizontal coordinates. The set of
\textit{complex} Fourier coefficients $\zeta_n$ are the components
of the eigenvector related to the linear stability problem and
determine the subharmonic time dependence. The spatial modes are
characterized by the wave vectors ${\bf k}_i$ , each carrying an
individual \textit{complex} amplitude $A_i$. These quantities are
determined by the nonlinearities of the problem. In principle the
${\bf k}_i$ can have any length and orientation but at onset the
relation $|{\bf k}_i|= k_c$ holds. The number N of participating
modes determines the degree of rotational symmetry of the pattern:
$N=1$ corresponds to lines, $N=2$ to squares, $N=3$ to hexagons or
triangles, etc. It can be shown \cite{douady89} that the $\zeta_n$
and $\zeta_{-n}$ are coupled in a way that $\zeta_n=\zeta_{-n}$ so
that heterodyning of right and left travelling waves always result
in \textit{standing} waves. Equation \ref{eqn1} then reads
\begin{multline}
h({\bf r}, t)= \sum_{i=1}^{N} {(
    |A_i| cos{ {\bf k}_i \cdot {\bf r}+\phi_i} }) \\
\times \sum_{n=0}^{+\infty}  {(|\zeta_n|
    cos{(n+1/2) \Omega t+\psi_n})}
\label{eqn3}
\end{multline}
The complex eigenvectors $\zeta_n$ can be calculated modulo a
constant factor and the ratio of the amplitudes
$|\zeta_n|/|\zeta_{n+l}|$ as well as the temporal phases $\psi_n$
can be compared with experimental data. They are obtained in the
following way: For each step in the driving amplitude a series of
snapshots of the surface state (Fig. \ref{squ}) is taken. The
primary pattern consists of squares and their formation is
governed by the nonlinearities of the problem and one of our goals
is to identify how far from onset the predictions from the linear
theory hold.

\begin{figure}
\includegraphics[ width=0.95\linewidth]{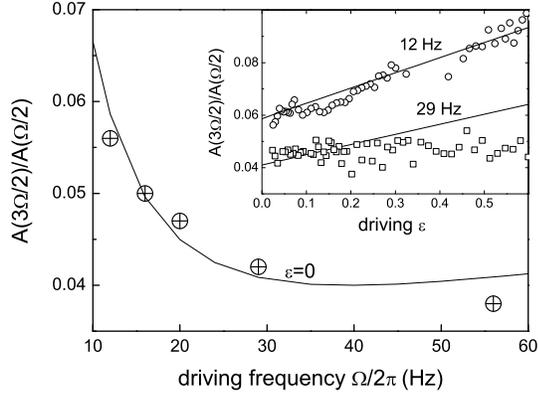}
 \caption{The ratio of the amplitudes
$A(\Omega3/2)/A(\Omega/2)$ of the ${\bf k}(10)$ mode at
$\varepsilon=0$ for different driving frequencies. The values are
extrapolated from measurements at $\varepsilon>0$ shown in the
inset: the  amplitude ratios at $\Omega/2\pi=12Hz$ and $29 Hz$ as
a function of the driving strength $\varepsilon$.} \label{lin2}
\end{figure}

An analysis of the Fourier transformation of the pictures yield
amplitudes $A(t)({\bf k}(ij))$ that are shown in Fig.
\ref{12hzsq1}. For the wave vectors ${\bf k}(ij)$ the nomenclature
from crystallography is used, e.g. ${\bf k}(10)$ and ${\bf k}(01)$
are the vectors that generate the simple unit cell of the square
pattern (Fig. \ref{squ}). The temporal evolution of the amplitude
of one of the critical modes $A(t){\bf k}(10)$ with $|{\bf
k}(10)|=\textit{k}_c$ (Fig. \ref{12hzsq1} is then again Fourier
transformed and a typical spectrum is shown in Fig.
\ref{12hzspec}a. These data are taken for all driving strengths
$\varepsilon$ (Fig. \ref{12hzbif1}a) and we always find the same
values for $A(t){\bf k}(10)$ and $A(t){\bf k}(01)$ within the
experimental resolution. In agreement with former investigations
\cite{Wernet01} in a system with a larger aspect ratio (container
size to wave length) our study reveals also that the fundamental
spatial mode $|{\bf k}(10)|=\textit{k}_c$ for all $\varepsilon$.
We can now extract the ratio of $A(\Omega3/2,{\bf
k}(10))/A(\Omega/2,{\bf k}(10))$ that is shown in the inset of
Fig. \ref{lin2}.
\begin{figure}
\includegraphics[ width=0.95\linewidth]{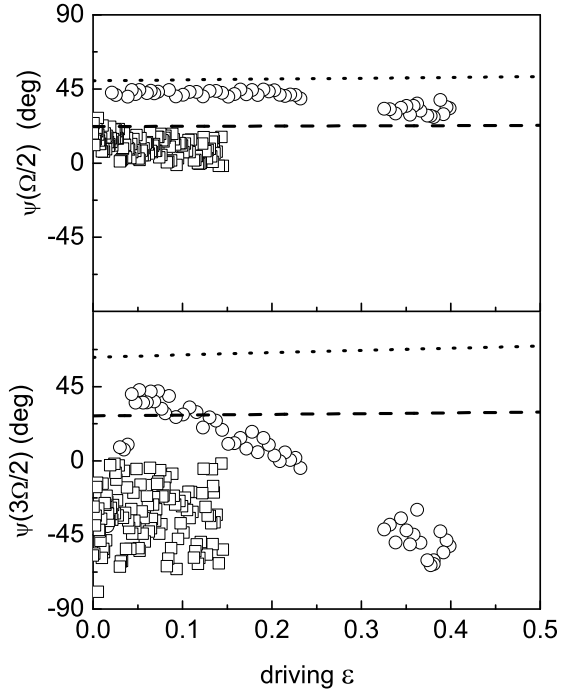}
\caption{The temporal phases $\psi$ ($\Omega/2)$ and
$\psi(3\Omega/2)$ of the ${\bf k}(10)$ mode for two driving
frequencies versus driving strength $\varepsilon$. The symbols
mark experimental, the lines theoretical data. Squares and broken
line: $\Omega/2\pi=57 Hz$. Circles and dotted line:
$\Omega/2\pi=12 Hz$; in the range $0.2<\varepsilon<0.28$ a
transition from squares to hexagons takes place and in this
disordered state an extraction of phases is not
possible.}\label{lin3}
\end{figure}
The contribution of higher harmonics  is in the order of $5$ to
$10\%$ and increases slightly with the driving strength. The
agreement between the experimental data and the linear theory is
again very good up to driving strength $\varepsilon
>0.5$, especially for lower driving frequencies. It is very
surprising that the agreement even holds up to secondary patterned
surface states, where a transition from a square to a hexagonal
state has been taken place and, as we show later, strong nonlinear
contributions participate in the dynamics of the system. The
experimental data show also that at driving strength as low as
$\varepsilon = 0.02$ the surface state consists of no measurable
higher spatial Fourier modes (see Fig. \ref{12hzbif1}) but of
higher temporal harmonics in perfect agreement with the linear
theory. This allows an extrapolation of $A(\Omega3/2,{\bf
k}(10))/A(\Omega/2,{\bf k}(10))$ to the neutral situation
$\varepsilon = 0$ for all driving frequencies $\Omega$ (Fig.
\ref{lin2}). The frequency ratio decreases first with increasing
frequency and has a minimum at $\Omega/(2 \pi)\approx 40 Hz$. This
characteristic shape reflects the amount of damping present in the
system. At low driving frequencies the ration between fill hight
and wave number is small. In this regime damping from the bottom,
that increases with decreasing frequency, is most significant. For
larger driving frequencies the damping from the bulk of the liquid
(a function increasing with the frequency) is the strongest
contribution. This behavior is also reflected in the critical
accelerations (compare with Fig. \ref{lin1}). The ratio
$A(\Omega5/2,{\bf k}(10))/A(\Omega/2,{\bf k}(10))$ has been
evaluated too, but the experimental resolution is not sufficient
here for a conclusive comparison between theory and experiment.

In the same way the temporal phases $\psi_n$ can be extracted from
the Fourier spectrum and once more a good agreement between the
theoretical predictions and experimental data is obtained, at
least for the fundamental $\Omega/2$ component. For the $3
\Omega/2$ component the scatter of the experimental data is very
large and  we find significant differences between experiment and
theory. But besides the large scatter we observe a pronounced
nonmonotonic behavior of the $\psi_{3/2\Omega}$ component at $57
Hz$ and this part of the spectrum seems to be governed by
nonlinear interactions.

\section*{4. The nonlinear surface state at $\Omega =12 Hz$}
\subsection {The square state}

\begin{figure}
\includegraphics[ width=0.95\linewidth]{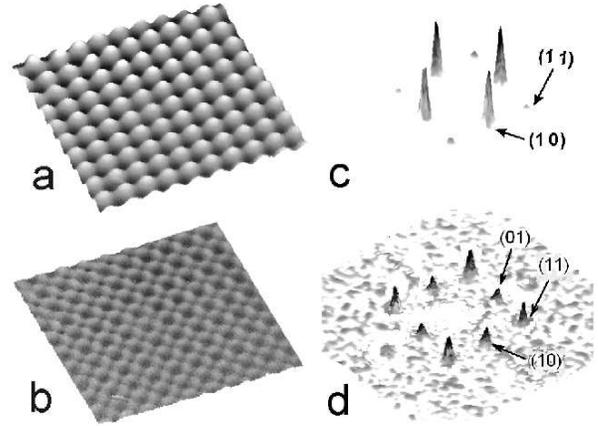}
\vspace{0.3cm} \caption{Snapshots of the surface state and the
power spectra at $\Omega/2\pi=12Hz$ and $\varepsilon=$0.17
($a_0=$30.0 m/s$^2$) for two different temporal phases a) at
maximum  and b) minimum surface elevation as indicated in Fig.
\ref{12hzsq1}}. \label{squ}
\end{figure}

\begin{figure}
\includegraphics[ width=0.95\linewidth]{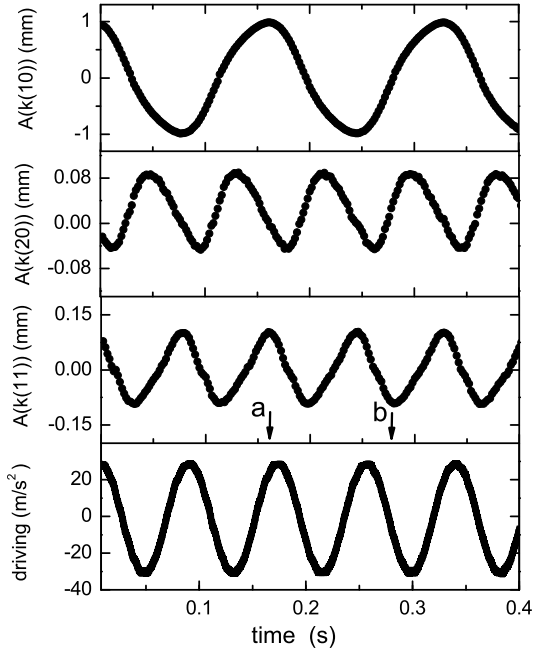}
\vspace{-0.3cm} \caption{The absolute amplitudes $A$ of different
spatial modes and the driving signal $a(t)$ in the square state at
$\Omega/2\pi=12Hz$ and $\varepsilon=$0.17 ($a_0=$30.0 m/s$^2$).
Please note that unlike in Ref. \cite{Kityk04} not the square root
of the power spectra but the amplitude A of a deformation $h({\bf
r},t)=A cos({\bf k}(ij) \cdot {\bf r})$ is shown.} \label{12hzsq1}
\end{figure}

\begin{figure}
\begin{flushleft}
\includegraphics[ width=1.05\linewidth]{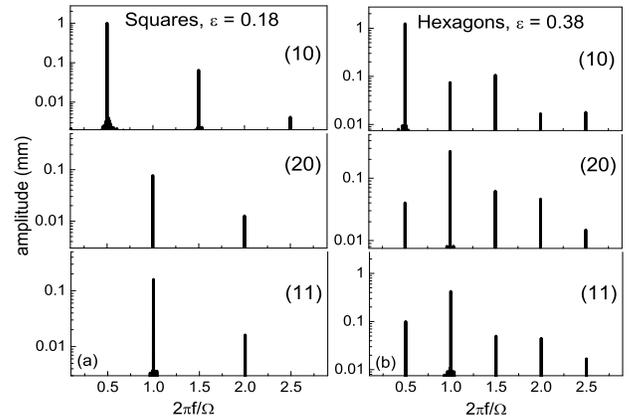}
\end{flushleft}
\vspace{-0.3cm} \caption{The temporal spectra of the amplitudes
$A({\bf k}(ij,\omega))$ of different spatial modes at
$\Omega/2\pi=12Hz$ and a driving strength: a) $\varepsilon$=0.18
in the square state and b) $\varepsilon=$0.38 in the hexagonal
state. } \label{12hzspec}
\end{figure}

\begin{figure}
\includegraphics[ width=0.52\linewidth]{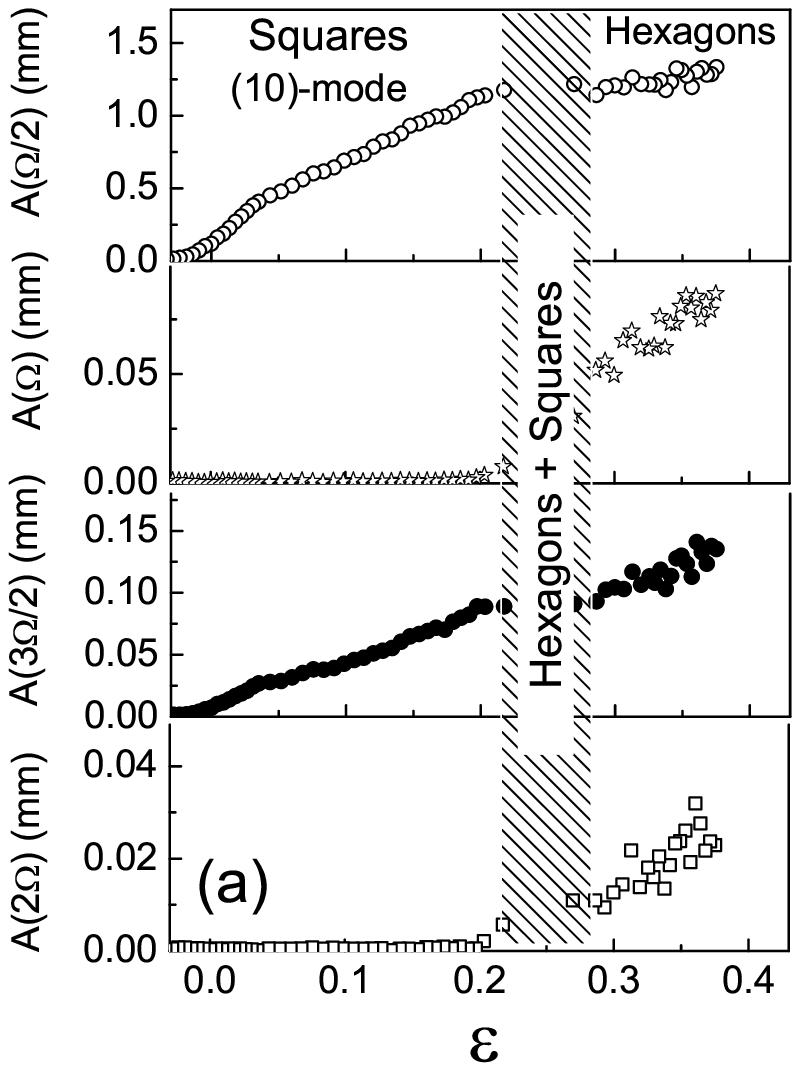}
\hspace{-0.6cm}
\includegraphics[ width=0.52\linewidth]{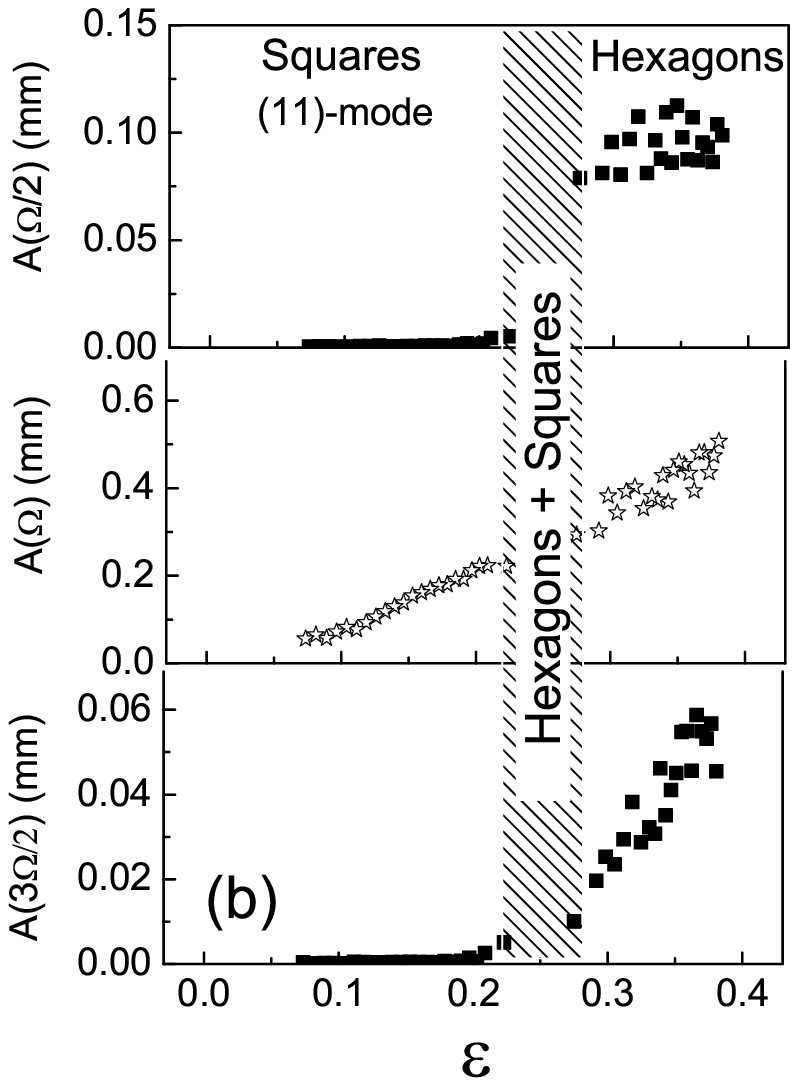}
\includegraphics[ width=0.78\linewidth]{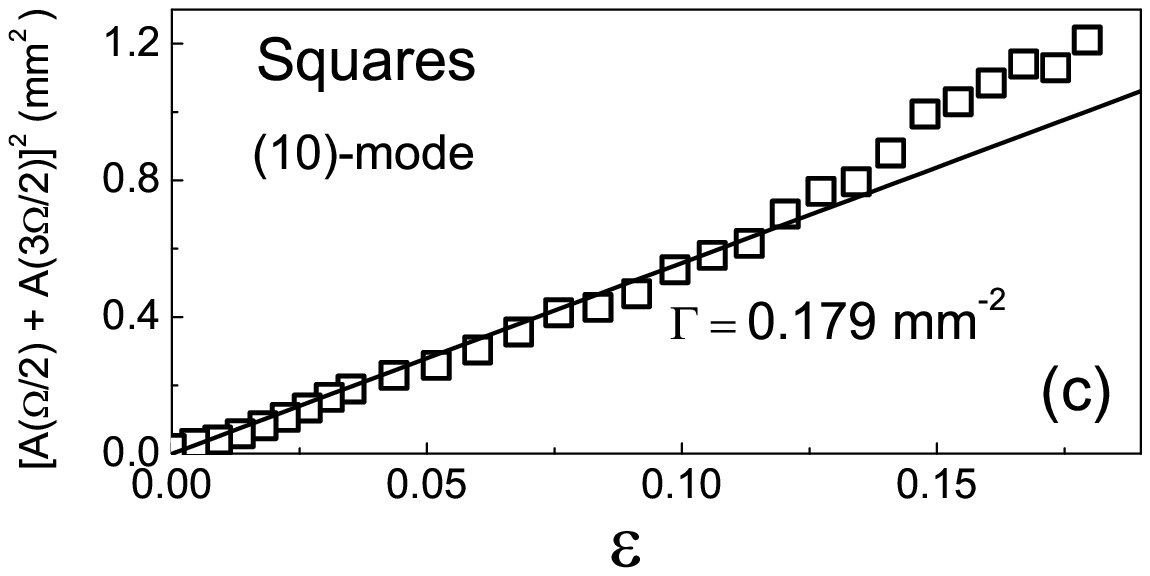}
\caption{The amplitudes $A(n/2\Omega)$ of the a) ${\bf k}(10)$ and
b) ${\bf k}(11)$ mode at $\Omega/2\pi=12Hz$ as a function of the
driving strength $\varepsilon$. c): the square of the sum of the
subharmonic components of the $A(10)$ mode versus $\varepsilon$.
As expected for a forward bifurcation the data can be linearly
fitted, at least up to driving strength of $\varepsilon \approx
0.1$ } \label{12hzbif1}
\end{figure}

The primary pattern near onset ($0<\varepsilon<0.28$) consists of
squares, shown in Fig. \ref{squ}. Their formation is determined by
the minimum of the Lyaponov functional of the according amplitude
equation of the critical modes \cite{cross93} and a quantitative
theoretical prediction of the expected pattern can be given by
inspection of the cubic coupling coefficient \cite{zhang96}. To
our knowledge, for a two liquid system there has not yet been an
attempt to calculate this coefficient, but squares are a common
pattern in free surface experiments with low viscous liquids. The
amplitude equations follow from a solvability condition of a
weakly nonlinear analysis of the underlying constitutive equation,
and its principal form is determined by the symmetries of the
system. For the subharmonic response one can write
\begin{equation}
\label{ampleqn1} \tau \partial_t A({\bf k}_i)= \epsilon  A({\bf
k}_i) - \sum_{j=1}^{N} \Gamma(\theta_{ij})|{A({\bf k}_j)}|^2
A({\bf k}_i),
\end{equation}

with $\tau$ the linear relaxation time and $\Gamma(\theta_{ij})$
the cubic coupling coefficient that depends on the angle
$\theta_{ij}$ between the modes ${\bf k}_j$ and ${\bf k}_j$ with
$|{\bf k}_{i,j}|=k_c$. The amplitudes $A({\bf k}_i)$ are modulated
with a subharmonic $(n+1/2)\Omega$ time spectrum given by the
$\zeta_n$ from the linear eigenvectors. Equation \ref{ampleqn1}
predicts a pitchfork bifurcation and in order to study this
scenario one has to extract the different temporal Fourier modes
of the measured $A({\bf k}_i,t)$ (Fig. \ref{12hzsq1}) first. The
result is shown in Figs. \ref{12hzspec} and \ref{12hzbif1}a. As
long as the pattern consists of squares there are no harmonic time
dependencies in the basic spatial modes to observe, but a
continued growth of $\Omega/2$ and $3\Omega/2$ contributions. The
$5\Omega/2$ contribution is very weak and only slightly larger
than the noise. The square of the sum of the amplitudes
$A_s=A(\Omega/2) + A(3\Omega/2)$ yields a straight line if plotted
versus the driving strength $\varepsilon$ (Fig. \ref{12hzbif1}c)
as one would expect for the case of a pitchfork bifurcation. From
the slope we can extract the cubic coupling coefficient
$A_s=\varepsilon/\Gamma(90^\circ)$, and we find
$\Gamma(90^\circ)=0.179 mm^{-2}$.

Now we can inspect the next higher harmonic spatial modes $A({\bf
k}(11))$ and $A({\bf k}(20))$. Their temporal evolution is shown
in Fig. \ref{12hzsq1}. Both modes  are a result of an interaction
of two fundamental modes, ${\bf k}(11)={\bf k}(10)+{\bf k}(01)$
and ${\bf k}(20)={\bf k}(10)+{\bf k}(10)$. Quadratic coupling does
not appear in the amplitude equations, but they are a natural
consequence of nonlinear spatial wave interaction and it is no
surprise that we find that they obey harmonic oscillations, shown
in Fig. \ref{12hzspec}a. The striking result of our analysis is
rather the constant offset that we find in the $A({\bf k}(11))$
and $A({\bf k}(20))$ spectrum (Fig. \ref{12hzsq1}). This means
that in addition to the oscillatory part, the interfacial profile
is also composed of contributions of constant deformations of the
form $h({\bf r}, t)= |A_i| cos{ {\bf k}_i \cdot {\bf r}}$. This
might surprisingly first, but please note that this does not
violate the mass conservation. Actually, it is a simple
consequence of the quadratic coupling of a \textit{real} standing
surface wave oscillation, $\Re (e^{i {\bf k}_i \cdot {\bf r}} e^{i
{\bf k}_i \cdot {\bf r}} e^{i\Omega/2}e^{-i\Omega/2})=cos{2 {\bf
k}_i \cdot {\bf r}} (1+cos{\Omega})$ (compare also Fig.
\ref{lines-ampl}c).

\begin{figure}
\includegraphics[ width=0.75\linewidth]{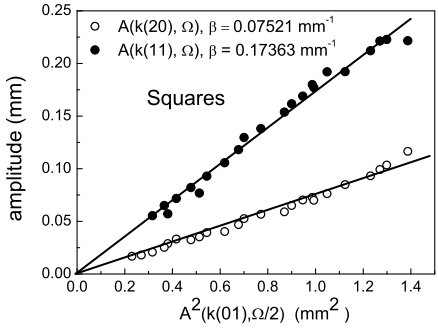}
\caption{The amplitudes $A({\bf k}(20))$ and $A({\bf k}(11))$
versus the square of the amplitude of the fundamental mode $A({\bf
k}(10))$ or the product of $A({\bf k}(10))$ and $A({\bf k}(01))$
respectively ($\Omega/2\pi=12Hz$). } \label{A1versA2}
\end{figure}

This quadratic coupling scheme can be verified by plotting $A({\bf
k}(20))$ and $A({\bf k}(11))$ versus the square of the amplitude
of the fundamental mode $A({\bf k}(10))$ or the product $A({\bf
k}(10))\times A({\bf k}(01))$ respectively. The data can be
perfectly reproduced by a linear fit. From the slope one gets the
strength of this nonlinear coupling and we do find the same values
for all frequencies $\Omega/2\pi = 16,20,29$ Hz where squares are
to be observed. Finally our Fourier analysis yields that the
imaginary part of the coupling scheme obeys the same resonance
conditions, and the spatial phase of the higher harmonic modes is
given by $\phi({\bf k}(20))=2\phi({\bf k}(10))$ and $\phi({\bf
k}(10))+\phi({\bf k}(01))$.

\subsection {The hexagonal state}
\begin{figure}
\includegraphics[ width=0.99\linewidth]{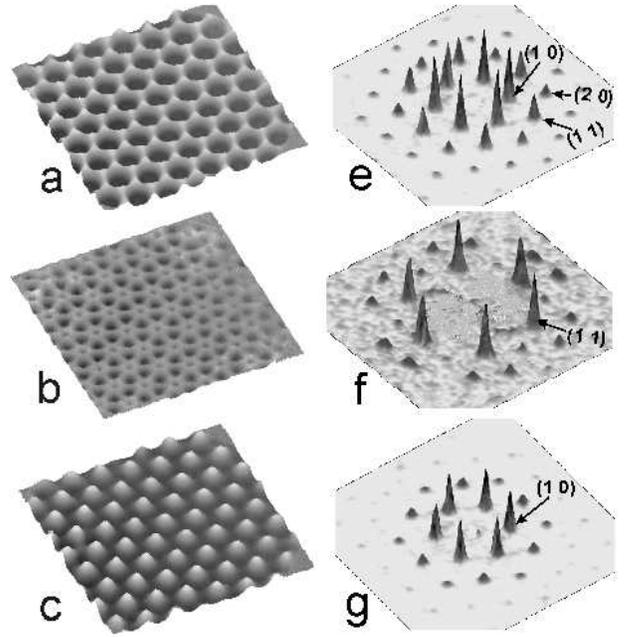}
\vspace{0.3cm} \caption{Snapshots of the surface state and the
Power spectra at $\Omega/2\pi=12Hz$ and $\varepsilon=$0.37
($a_0=$39.3 m/s$^2$) for three different temporal phases. a) down
hexagons, b) minimal surface elevation, c) up hexagons. See Refs.
\cite{wagner00,Kityk04} for further explanations on the switch
from up to down hexagons in the Faraday-Experiment. } \label{hex}
\end{figure}

\begin{figure}
\includegraphics[ width=0.5\linewidth]{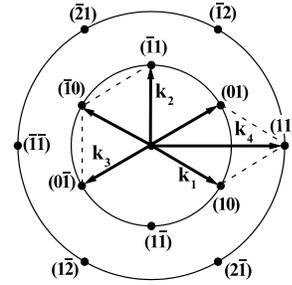}
 \caption{Vector diagram of the interacting modes
for the hexagonal surface state. } \label{kvectorshex}
\end{figure}

In the range ($0.20<\varepsilon<0.28$) the pattern becomes
disordered and transforms at higher driving strength to a
hexagonal state (see Fig. \ref{hex}) that consists of three
fundamental spatial Fourier modes ${\bf k}_{1,2,3}$. But please
note that for the construction of the crystallographic simple unit
cell two vectors ${\bf k}(10)={\bf k}_1$ and ${\bf
k}(\bar{1}1)={\bf k}_2$ are sufficient (${\bf k}(\bar{1}1)+{\bf
k}(10)={\bf k}(0\bar{1})={\bf k}_3$, as indicated in Fig.
\ref{kvectorshex}).

\begin{figure}
\includegraphics[ width=0.99\linewidth]{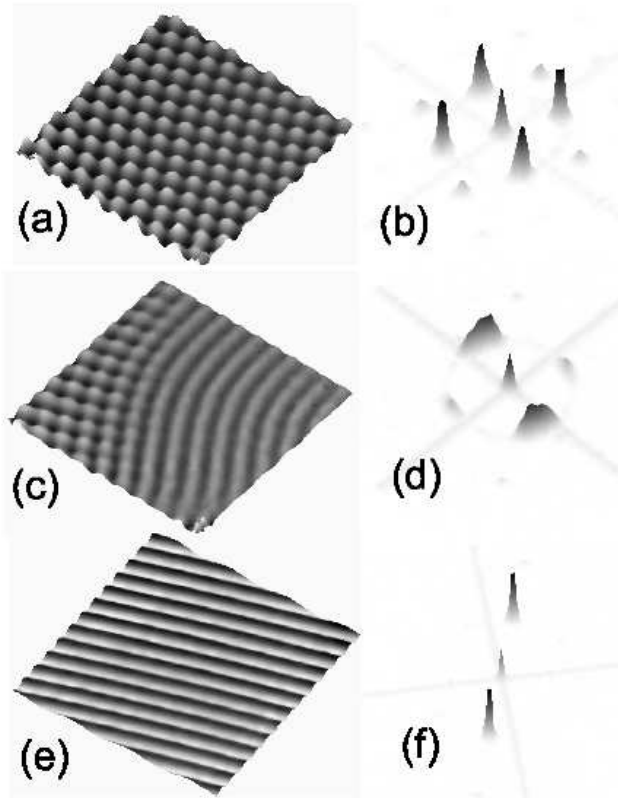}
\vspace{0.3cm} \caption{Snapshots of the surface state and the
power spectra at a,b) $\Omega/2\pi=$20Hz, $\varepsilon=$0.6
($a_0=$54.4 m/s$^2$) c,d) $\Omega/2\pi=$20Hz, $\varepsilon=$0.08
($a_0=$36.7 m/s$^2$) e,f) $\Omega/2\pi=$57Hz, $\varepsilon=$0.11
($a_0=$116.3 m/s$^2$). } \label{sqline}
\end{figure}

\begin{figure}
\includegraphics[ width=0.75\linewidth]{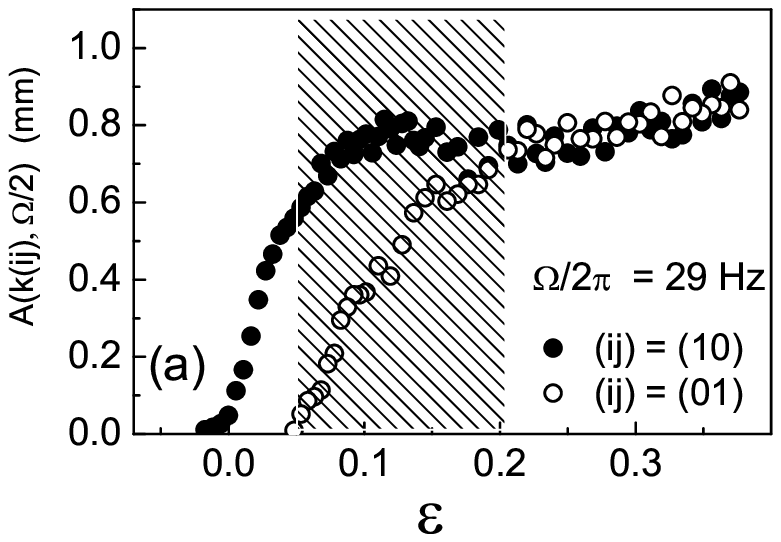}
\includegraphics[ width=0.75\linewidth]{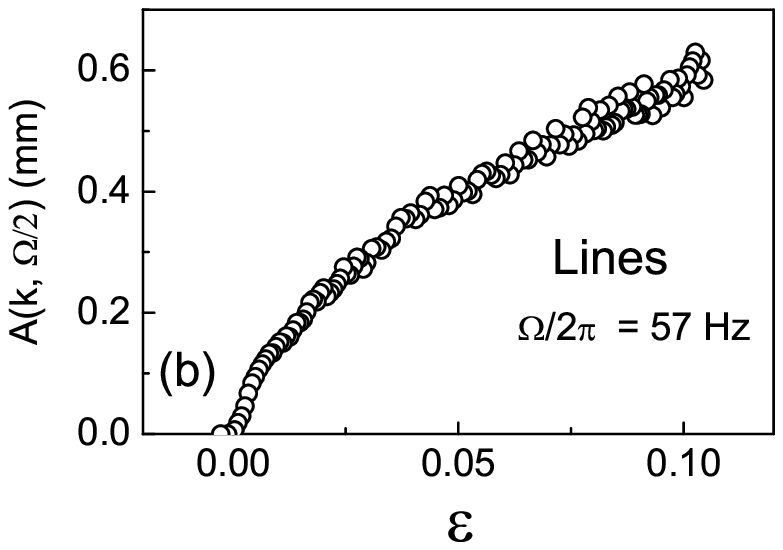}
\vspace{0.3cm} \caption{The amplitudes $A(\Omega/2)$ of the a)
${\bf k}(10)$ and ${\bf k}(01)$ mode at $\Omega/2\pi=29Hz$ and b)
${\bf k}(1)$ mode at $\Omega/2\pi=57Hz$ as a function of the
driving strength $\varepsilon$. The shaded region in a) indicates
the crossed rolls state.} \label{sqlinebif}
\end{figure}

An analysis of the temporal behavior of the amplitudes $A({\bf
k}(ij))$ of the spatial modes reveals, besides the the striking
offset with the according constant spatial sinusoidal surface
deformation, both harmonic and subharmonic time dependencies (see
Fig. \ref{12hzspec}). While harmonic ($n\Omega$) temporal
contributions in the higher spatial harmonics ${\bf k}(20,11)$
appear in a similar manner to that of the square pattern, the
resonance between ${\bf k}(10)+{\bf k}(0\bar{1})={\bf
k}(1\bar{1})$ results in harmonic ($n\Omega$) contributions in the
critical mode $|{\bf k}(1\bar{1})|=\textit{k}_c$. Consequently the
$n \Omega$ contributions couple with $(n+1/2) \Omega$
contributions back into the spectrum of the ${\bf k}(20,11)$ modes
and result in subharmonic contributions. Harmonic contributions do
not appear in the temporal spectra of the linear unstable modes
$\textit{k}_c$ and quadratic interactions of ($n \Omega/2$)
components do not appear in the amplitude equations. Nevertheless
the hexagonal state allows for a \textit{spatial} resonance
between linear unstable modes. In other words, this means that -
within the framework of the weakly nonlinear approximation - we
have here the interesting case where the system has a broken
temporal symmetry that is driven by spatial resonances. It is not
to be observed at any point in the quadratic state, where spatial
resonances between linear unstable modes are forbidden too. This
particular violation of the weakly nonlinear resonance conditions
can best be seen in Fig. \ref{12hzbif1} where clearly the
amplitudes of the $\Omega$ components of the $|{\bf
k(10)}|=\textit{k}_c$ mode grow from zero at the transition point
from squares to hexagons, and similarly the $\Omega/2$ components
of the $|{\bf k(11)}|=2\textit{k}_c$ modes.

\section*{5. Pattern dynamics at $\Omega/2\pi> 12 Hz$}
\begin{figure}
\includegraphics[ width=0.95\linewidth]{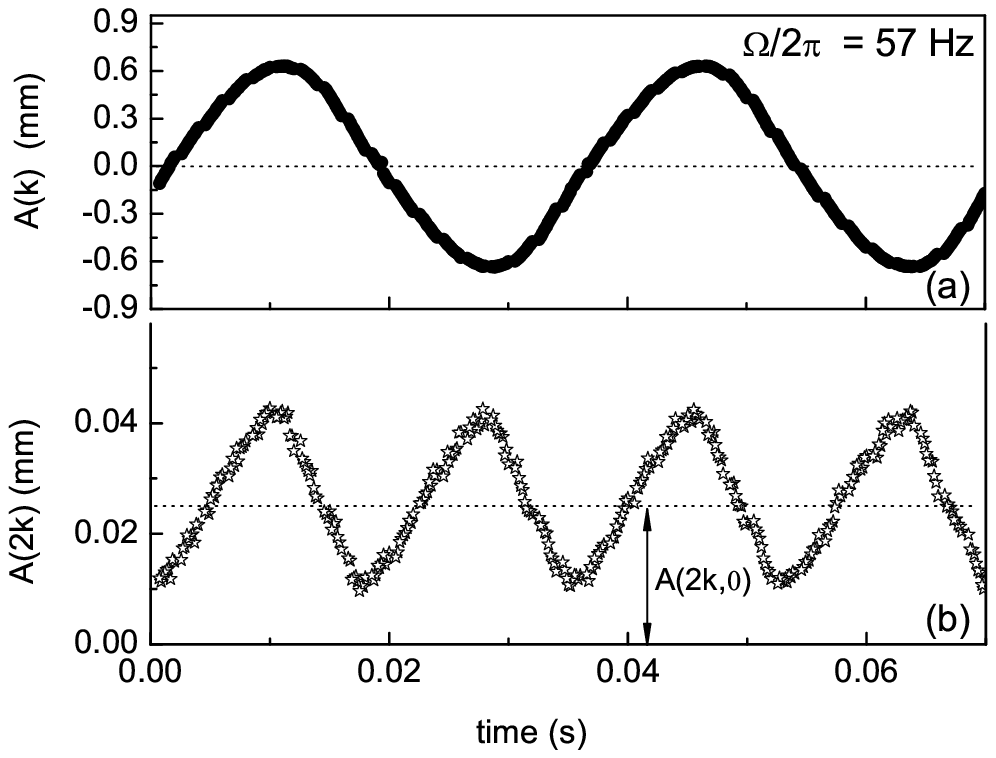}
\includegraphics[ width=0.95\linewidth]{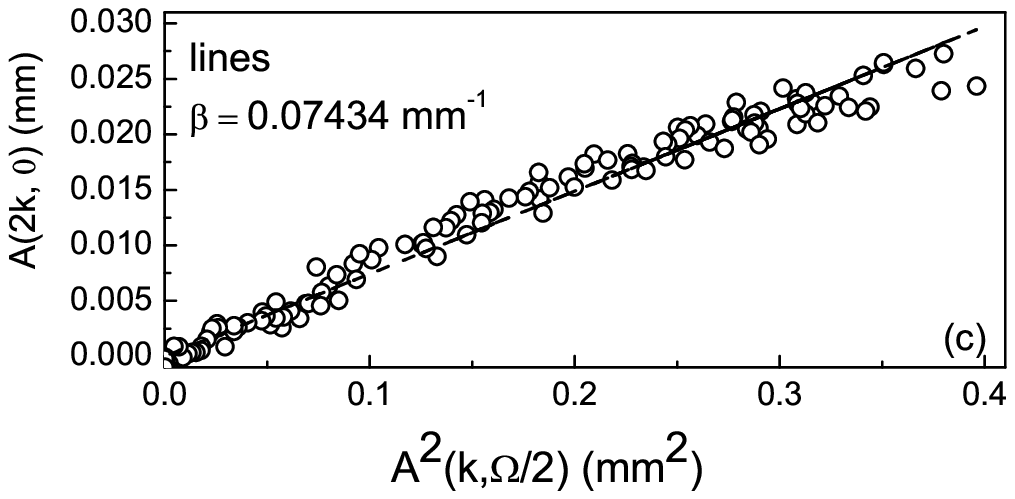}
\vspace{0.3cm} \caption{a,b): The absolute amplitudes $A$ of
different spatial modes in the line state at $\Omega/2\pi=$57Hz
and $\varepsilon=$0.11 ($a_0=$116.3 m/s$^2$). In b) the temporal
constant offset $A(2k,0)$ is indicated by the dotted line. c) The
amplitude $A(2k,0)$ versus $A^2(k,\Omega/2)$} \label{lines-ampl}
\end{figure}
The pattern dynamics at driving frequencies $\Omega/2\pi> 12 Hz$
are characterized by a transition to lines. At $\Omega/2\pi= 16
Hz$ the pattern still consists only of squares, while at
$\Omega/2\pi= 20$ and $29 Hz$ the primary pattern consists of
(slightly distorted) lines (Fig. \ref{sqline}c,d). At higher
driving strengths $\varepsilon$, a second Fourier mode
perpendicular to the first one starts to grow (Fig.
\ref{sqlinebif}a) and the pattern evolves to a square state (Fig.
\ref{sqline}a,b). For $\Omega/2\pi= 57 Hz$ a pure line state is
stable for all driving strengths (Fig. \ref{sqline}e,f and
\ref{sqlinebif}b). The pronounced constant offset in the $A(2{\bf
k},t)$ (Fig. \ref{lines-ampl}b) mode is now larger than the
temporal oscillation period and $A(2{\bf k},t)$ never crosses the
zero line. Similar like for the $A({\bf k} 20,\Omega)$ or $A({\bf
k} 11,\Omega)$ modes in the squares state this quadratic coupling
scheme holds also for the zero frequency modes as shown in (Fig.
\ref{lines-ampl}c).

\section*{6. Conclusion}
We have demonstrated a new technique to measure quantitatively the
spatio-temporal Fourier spectrum of Faraday waves on a two liquid
interface. With this technique it is now possible to test
theoretical predictions, especially those from numerical
simulations. To our knowledge there are still no full Navier
Stokes numerical simulation of the 3D problem and quantitative
tests for future work are most important. In this sense we would
like to encourage such attempts. But with our technique we are
also able to verify known predictions from the linear stability
analysis and we find good agreement up to high driving strength of
$\varepsilon \approx 0.5$. In the nonlinear state the most
pronounced result is the identification of strong temporal
constant sinusodial surface deformations in the spectrum. And with
our possibility to access any Fourier component separately we can
identify several resonance mechanisms, including an interesting
case of a temporal resonance violation by use of spatial
resonances.

\begin{acknowledgments}
This work was supported by the German Science Foundation project
Mu 912.
\end{acknowledgments}

\end{document}